# Beating the average: how to generate profit by exploiting the inefficiencies of soccer betting

Ralph Stömmer *

(Working paper, version 2, November 2, 2026)

**Abstract**

In economy, markets are denoted as efficient when it is impossible to systematically generate profits which outperform the average. In the past years, the concept has been tested in other domains such as the growing sports betting market. Surprisingly, despite its large size and its level of maturity, sports betting shows traits of inefficiency. The anomalies indicate the existence of strategies which shift betting from a game of chance towards a game of skill. This article shows an example for an inefficiency detected in the German soccer betting TOTO 13er Wette, which is operated by state-run lottery agencies. Gamblers have to guess the outcome (win, draw, loss) of 13 soccer matches listed on a lottery tip. Applying stochastic methods, a recipe is presented to determine hit rates for single match outcomes. More important, the recipe provides the number of lottery tips required to achieve a specific number of strikes (number of correct match forecasts per lottery tip) for any given level of safety. An approximation is derived to cope with large numbers in hypergeometric distributions, valid under certain constraints. Overall, the strategy does lead to returns exceeding the aggregated lottery fees, resulting in moderate, but consistent profits. It is briefly discussed if lessons learned from soccer betting can be transferred back to financial markets, because gamblers and retail investors face similar challenges and opportunities.

---

* Private researcher, e-mail: ralph.stoemmer@trajectorix.de

# 1. Introduction

## 1.1. The efficient market hypothesis

Economists denote markets as efficient, when it is impossible to systematically generate profits which outperform the average. Stock markets often serve as an example: any short term excess gains achieved by individual shareholders tend to be interpreted as statistical outliers, rather ascribed to chance and to lucky timing than to talent and to expertise. It is assumed that markets featuring countless participants in competition to each other, information available to all, numerous trades, and inherent randomness balance towards a state of equilibrium characterized by efficiency. Any news about single shares and the market would be quickly digested and reflected in asset prices without delay. Participants´ aggregated knowledge would repeatedly outperform pundit recommendations and economists forecasts, and anomalies such as over- and underreaction to available information would level out [1, 2, 3, 4, 5]. The observation that even professionals would be unable to outperform the collective judgement of the crowd is generally paraphrased with “long term, one can´t beat the average”.

## 1.2. The inefficiency of sport betting markets

Market efficiency has been originally formulated for financial markets, but applies in a natural way to betting markets where asset prices are replaced by betting odds. Accordingly, the efficiency hypothesis has been tested in the sports betting market, which over the past decades has become a global industry attracting companies and gamblers worldwide. The European Gaming and Betting Association reports that just in Europe more than 10 billion € annual revenue is generated from regulated online sports betting [6]. Gamblers can easily shop around between many betting agencies and bookmakers. Contrasting financial markets, prices are not dictated by supply and demand, but set by the betting agencies through fixed wagers and betting odds ratios [7]. The expected profit for the bettor is shifted by default into the negative with betting odds, fees and wagers, which indicates that long term, gamblers are expected to lose. The business model of betting agencies makes sure that any positive returns for gamblers do remain rare events, ascribed to luck only. Thaler et al. [8] once denoted betting markets as weak efficient, when no strategies exist to generate positive returns. Surprisingly, despite its large size and its level of maturity, sports betting shows traits of inefficiency [9, 10, 11, 12, 13]. In consequence, sports betting has been tackled applying heavy ordnance to gain profit, with artificial intelligence constituting the latest trend [14]. This work shows an example for an inefficiency detected in the German soccer betting TOTO 13er Wette, which is operated by state-run lottery agencies. Gamblers have to guess the outcome of 13 soccer matches listed on a lottery tip. Applying stochastic methods, a recipe is presented to determine tendencies for single match outcomes. More important, the recipe provides the number of lottery tips required to achieve a specific number of strikes (number of correct match forecasts per lottery tip) for a given level of safety. Overall, the strategy does lead to expected returns exceeding the aggregated lottery fees, resulting in consistent profits. Despite this inefficiency, generating profits high enough to earn one´s living turns out to be a challenge for other reasons: unforeseen transaction efforts and legal betting regulations imposing financial limits.

## 2. The German soccer betting TOTO 13er Wette (Normalschein)

### 2.1. Background

The soccer betting TOTO 13er Wette, in its basic version named Normalschein, allows to place bets on the outcome of soccer matches. Revenue is distributed to lucky gamblers who have correctly forecasted at least 10 out of 13 matches. The betting is operated weekly by state-run lottery agencies. Germany consists of 16 federal states, and each one individually assumes responsibility for issuing lottery vouchers, collecting the filled out lottery tips including fees from the individual gamblers, and rewarding the lucky winners. At the beginning of each week, 13 soccer matches, due at the weekend, are issued. The majority of matches are taken from the German premier soccer league Erste Bundesliga, supplemented by matches from the German second league Zweite Bundesliga, and foreign matches taken from any of the European premier leagues (for instance England, Italy, or Spain). The 13 soccer matches are listed in a column denoted as tip, and gamblers have to guess the outcome (win, draw, loss) for each single match. A home team win is denoted as number 1, a draw as number 0, and the away team win, which corresponds to the home team loss, is denoted as number 2. Figure 1a shows a copy of a lottery voucher. It is filled out and submitted to a local lottery shop, which acts as receiving office for the state-run lottery agency. Alternatively, the voucher can be filled out and directly submitted online to the state-run lottery agency, see sample in figure 1b from Lotto Bayern, available at www.lotto-bayern.de/toto/13er-

wette/normalschein. In both cases, offline as well as online, gamblers are provided with 12 tips, which allows for various combinations of possible match outcomes. The format of the vouchers, the lottery fees, and the way of the transaction of the bets to the lottery agencies are dependent on the preferred design of the federal states. In general, filling out a single tip with 13 matches costs 0,50€, which is equal in all federal states at the time of writing of this work. Federal states differ in the way of transaction, which complicates matters, as will be discussed in the results chapter. At the time of writing, just 8 out of 16 federal lottery agencies allow to submit the filled out voucher of the TOTO 13er Wette online.

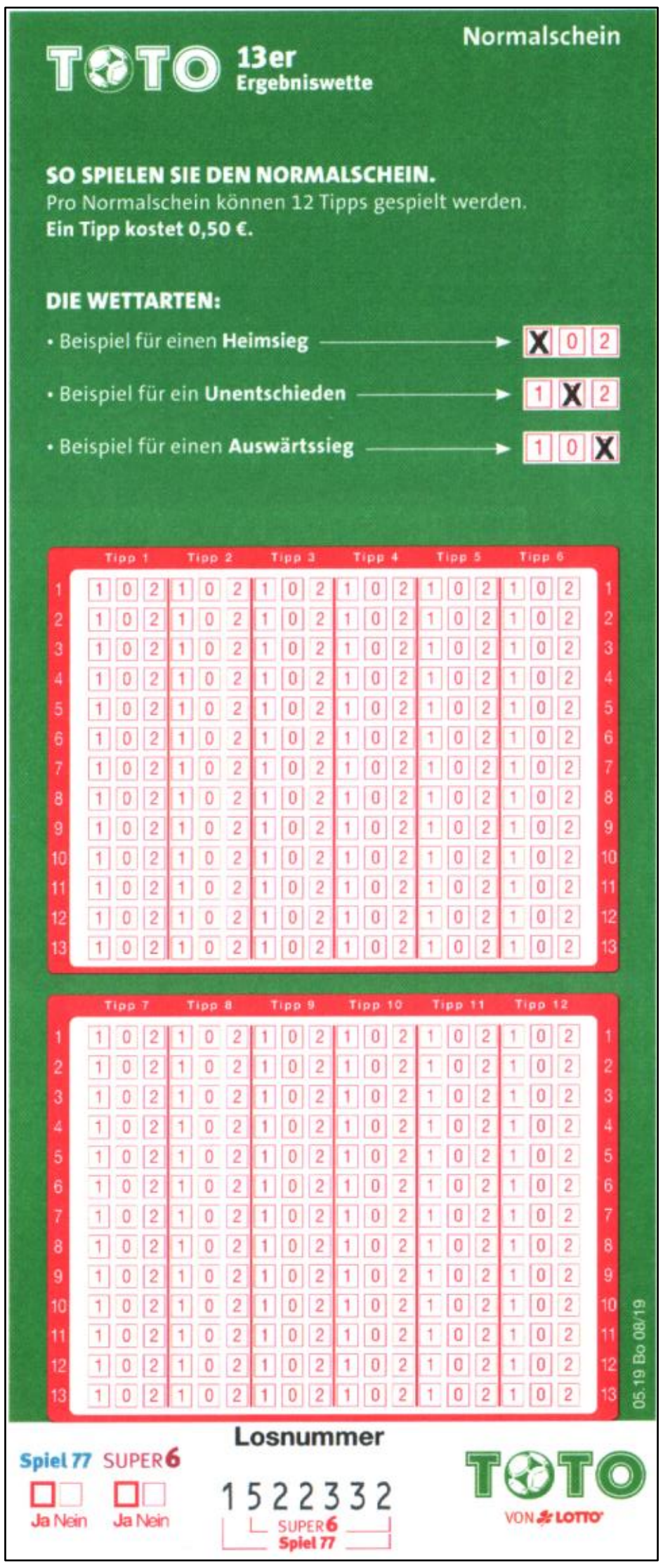

Figure 1a: Paper voucher from a local lottery shop, containing 12 tips. Each tip is comprised of 13 matches. Home win (1), draw (0), or away win (2) have to be checked with ball pen.

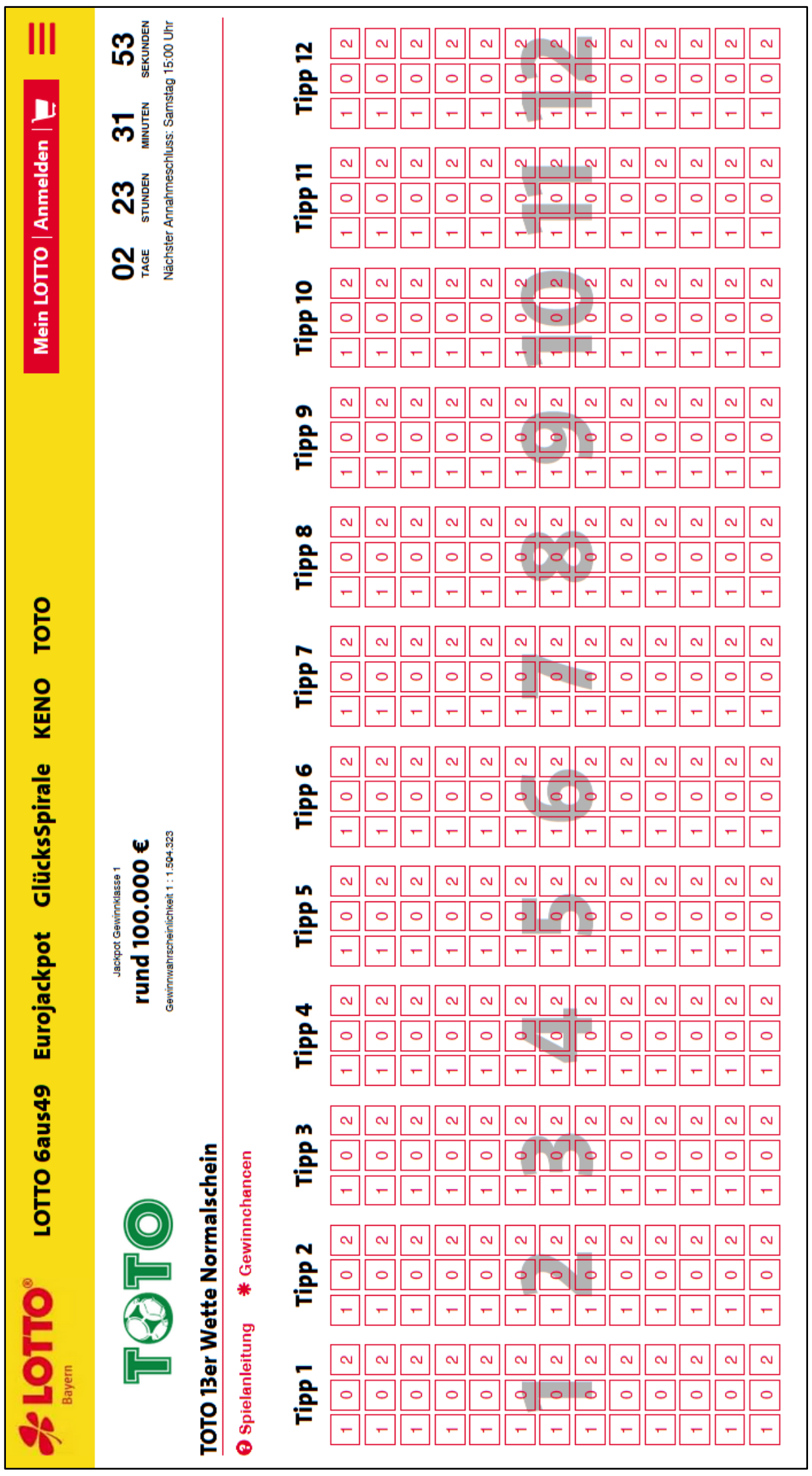

Figure 1b: Online voucher from Lotto Bayern, containing 12 tips. Each tip is comprised of 13 matches. Home win (1), draw (0), or away win (2) are filled out online. It is necessary to open an online account first.

Additional fees for submitting a single voucher online vary in between 0,20€ to 0,70€. Gamblers who want to file more than 12 tips have to submit additional vouchers. For the case the gambler does submit just 1 tip on a voucher online at the lottery agency Lotto Bayern, his total fee amounts to 0,50€ (1 tip) + 0,50€ (1 voucher) = 1,0€. If the gambler submits 20 tips, for which he requires 2 vouchers, the total fee amounts to 10€ (20 tips) + 1€ (2 vouchers) = 11€. The actual match plan of soccer matches due at the weekend is available in lottery shops, in gambler magazines or published online on the web pages of lottery agencies. Figure 2 provides an example for a partially filled out online voucher, with the outcome guesses of 13 matches in the column of the 1st tip. The example is created for demonstration purposes only. The first column lists the home teams for each of the 13 matches, the second column the away teams. In the third column named tendency, probabilities for the match outcome are provided by the lottery agency as extra bonus.

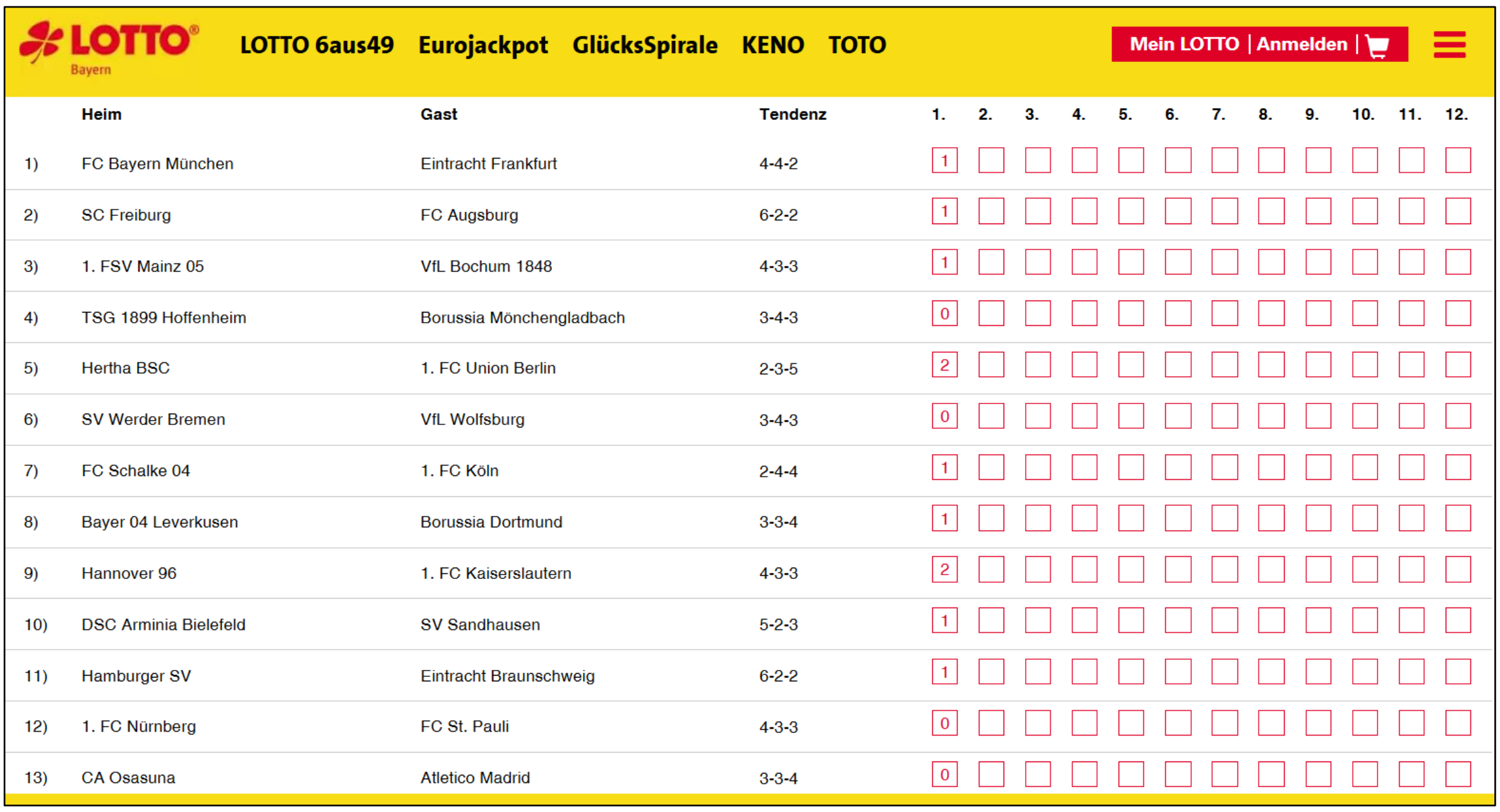

| | Heim | Gast | Tendenz | 1. | 2. | 3. | 4. | 5. | 6. | 7. | 8. | 9. | 10. | 11. | 12. |
|---|---|---|---|---|---|---|---|---|---|---|---|---|---|---|---|
| 1) | FC Bayern München | Eintracht Frankfurt | 4-4-2 | 1 | | | | | | | | | | | |
| 2) | SC Freiburg | FC Augsburg | 6-2-2 | 1 | | | | | | | | | | | |
| 3) | 1. FSV Mainz 05 | VfL Bochum 1848 | 4-3-3 | 1 | | | | | | | | | | | |
| 4) | TSG 1899 Hoffenheim | Borussia Mönchengladbach | 3-4-3 | 0 | | | | | | | | | | | |
| 5) | Hertha BSC | 1. FC Union Berlin | 2-3-5 | 2 | | | | | | | | | | | |
| 6) | SV Werder Bremen | VfL Wolfsburg | 3-4-3 | 0 | | | | | | | | | | | |
| 7) | FC Schalke 04 | 1. FC Köln | 2-4-4 | 1 | | | | | | | | | | | |
| 8) | Bayer 04 Leverkusen | Borussia Dortmund | 3-3-4 | 1 | | | | | | | | | | | |
| 9) | Hannover 96 | 1. FC Kaiserslautern | 4-3-3 | 2 | | | | | | | | | | | |
| 10) | DSC Arminia Bielefeld | SV Sandhausen | 5-2-3 | 1 | | | | | | | | | | | |
| 11) | Hamburger SV | Eintracht Braunschweig | 6-2-2 | 1 | | | | | | | | | | | |
| 12) | 1. FC Nürnberg | FC St. Pauli | 4-3-3 | 0 | | | | | | | | | | | |
| 13) | CA Osasuna | Atletico Madrid | 3-3-4 | 0 | | | | | | | | | | | |

**Figure 2**: Partially filled out online voucher, with the outcome guesses of 13 matches in the column of the 1st tip. The first column lists the home teams, the second column the away teams. In a third column named tendency, probabilities for the match outcome are provided.

The model applied to calculate the probabilities of the match outcomes is kept confidential by the lottery agencies. The number sequence 4-4-2 in the 1st row, home team FC Bayern München versus away team Eintracht Frankfurt, does mean: 40% probability that home team FC Bayern München will win the match, 40% probability for a draw, and 20% probability that away team Eintracht Frankfurt will win the match (corresponds to the loss of the home team FC Bayern München). There is no need for gamblers to adhere to the tendencies. If the gambler is confident that the home team FC Bayern München will win, the number 1 is put into the respective box in the tip columns on the right hand side in figure 2. For instance in the 9th row, the gambler deviates from the tendency. The number 2 in the respective box in the tip column indicates that the gambler rather believes in a win of the away team FC Kaiserslautern, in contrast to the provided 4-3-3 tendency, which indicates a 40% probability of the home team Hannover 96 to win the match.

## 2.2. Basic probabilities and crucial number of strikes per lottery tip

The betting agency distributes returns when at least 10 matches out of 13 matches of the TOTO 13er Wette are correctly forecasted by the gambler. Assuming that there is no prior information about the soccer teams, forecasting becomes purely random. Win, draw and loss constitute 3 possible match outcomes, so the hit rate p to correctly forecast the outcome of a single match reduces to $p = 1/3$. With even chances, the probability to correctly forecast all 13 match results on a single lottery tip calculates to $P(13) = p^{13} = 1/1.594.323$. Correctly forecasted match results on a tip will be named strikes. Table 1 shows the probabilities $P(x)$ for $x = 10, 11, 12, 13$ strikes per tip, supplemented with the theoretical and the expected return. The 2nd column lists the quantity $M_x$, which is the number of combinations with x strikes. For instance, $M_{11} = 312$ indicates that there are 312 different combinations to achieve 11 strikes per tip, in a total amount of $N = 1.594.323$ combinations. The theoretical return, calculated for a representative pool, just serves as a rough guideline for bettors.

| x strikes per tip | Mx combinations with x strikes in N combinations | Probability P(x) to achieve x strikes per tip | Theoretical return* | Expected return |
|---|---|---|---|---|
| 10 | 2.288 | 1 / 697 | 52 € | 49 € |
| 11 | 312 | 1 / 5.110 | 306 € | 434 € |
| 12 | 26 | 1 / 61.320 | 3.679 € | 6.449 € |
| 13 | 1 | 1 / 1.594.323 | 167.404 € | 107.813 € |

**Table 1**: Basic figures for even chances, hit rate $p = 1/3$. $P(x)$ is the probability for achieving x strikes per lottery tip. $M_x$ denotes the number of combinations with x strikes in a total amount of $N = 1.594.323$ combinations. The theoretical return is given for reference, the expected return is relevant for further calculations. All quantities in € are rounded up to integer numbers. (*published in TOTO 13er Wette handout available in local lottery shops).

The actual return of the TOTO 13er Wette depends on the total pool, the number of gamblers participating, and the number of lucky winners achieving the relevant number of strikes. The expected return in the last column, which will be used for further calculations in this work, represents the average weekly return for each of the relevant strikes per tip, taking into account all weekly returns of the past 1,5 years.

## 3. Mathematical model

### 3.1. Probability for x strikes on a single tip: the binomial distribution

The probability P(x) to achieve x strikes on a single tip is given by the binomial distribution

$$P(x) = \binom{y}{x} p^{x} (1-p)^{y-x} \quad . \tag{1}$$

The hit rate p is the probability to correctly guess the outcome of a soccer match. When nothing is known about the strength of soccer teams, as mentioned already in the previous chapter 2.2, $p = 1/3$. The figures published on the soccer betting TOTO 13er Wette all refer to $p = 1/3$. y is the number of soccer matches on a single tip. With $y = 13$ soccer matches for the TOTO 13er Wette, equation 1 yields the probability values P(x) for x = 10, 11, 12, 13 strikes per tip listed in table 1. The full range of strikes would cover x = 0 - 13, but for x = 0 - 9, the betting agency does not distribute any returns.

Providing the probability values as fractions of integers allows to easily derive the number N of all combinations, making use of the highest possible number of strikes. 13 strikes appear in a single combination only, $M_{13} = 1$. Equation 1 gives $P(13) = p^{13}$. P(13) can be written as fraction $M_{13}/N$, which expresses the basic Laplace definition of probability. It follows that $N = (1/p)^{13} = 1.594.323$. In order to be 100% certain to achieve 13 strikes, it is required to fill out 1.594.323 tips, each tip with a unique combination (no doubles) to cover all possible results. According to the fees provided as examples in previous chapter 2.1, the total fee for tips and vouchers would amount to 863.592€, which by far exceeds the expected return for 13 strikes, which according to table 1 is 107.813€. 100% safety to win the soccer betting with 13 strikes seems achievable only with tremendous costs and efforts and a negative business case for the gambler. The question is, if a method can be developed to determine the number of required tips, which allows to achieve a specific number of strikes for a given level of safety.

### 3.2. Probability to draw exactly $k_x$ combinations with x strikes in n tips

A filled out tip can be imagined as a colored ball in the urn model. $M_x$ provides the number of combinations with x strikes, which corresponds to $M_x$ balls with the same color x. The basic condition is

$$\sum_{x=0}^{13} M_x = N \ . \qquad (2)$$

Within the framework of the urn model, each x = 0 - 13 can be interpreted as a specific color in a jar containing N balls with 14 different colors. Q($k_x$ in n) defines the probability to draw exactly $k_x$ combinations with x strikes in n tips, it is given by the hypergeometric distribution

$$Q(k_x \text{ in } n) = \frac{\binom{M_x}{k_x}\binom{N-M_x}{n-k_x}}{\binom{N}{n}} \ . \qquad (3)$$

Equations 1 and 3 are related to each other. The probability Q($k_x$ = 1 in n = 1), which is the probability to draw exactly one combination with x strikes on a single tip, calculates to

$$Q(k_x = 1 \text{ in } n = 1) = \frac{\binom{M_x}{1}\binom{N-M_x}{0}}{\binom{N}{1}} = \frac{M_x}{N} \ . \qquad (4)$$

The fraction $M_x/N$ on the right hand side in equation 4 equals P(x) in equation 1, a relation briefly introduced for 13 strikes in chapter 3.1:

$$\frac{M_x}{N} = P(x) \ . \qquad (5)$$

Equation 5 allows to derive $M_{10}$, $M_{11}$, $M_{12}$, and $M_{13}$ listed in table 1. Equation 3 provides the obvious, that one has to draw all possible combinations in order to be 100% safe to cover the combination with 13 strikes. With $M_{13}$ = 1 and n = N equation 3 yields

$$Q(k_{13} = 1 \text{ in } n = N) = \frac{\binom{M_{13}}{1}\binom{N-M_{13}}{N-1}}{\binom{N}{N}} = \binom{1}{1}\binom{N-1}{N-1} = 1 \ . \qquad (6)$$

Q($k_x$ in n) will later serve to define a level of safety for drawing a minimum number of combinations with x strikes in n tips. The multivariate hypergeometric distribution provides the probability to draw $k_{10}$, $k_{11}$, $k_{12}$, $k_{13}$ combinations with x = 10, 11, 12, 13 strikes in n tips:

$$Q(k_{10}, k_{11}, k_{12}, k_{13} \text{ in } n) = \frac{\binom{M_{10}}{k_{10}}\binom{M_{11}}{k_{11}}\binom{M_{12}}{k_{12}}\binom{M_{13}}{k_{13}}\binom{N-M_{10}-M_{11}-M_{12}-M_{13}}{n-k_{10}-k_{11}-k_{12}-k_{13}}}{\binom{N}{n}} \ . \qquad (7)$$

A gambler, who wants to determine the number n of tips required to achieve $k_x$ combinations with x strikes with a certain probability, for instance Q($k_x$ in n) = 99%, is faced with two challenges. First of all, the number n of required tips is contained implicitly in equation 7, it can´t be solved in a closed form to yield an explicit expression n = function(Q, $M_x$, $k_x$, N) for x = 10, 11, 12, 13. Instead, the gambler can stay with equation 7 and start computing Q($k_x$ in n) as a function of n. As a final step, the gambler would choose the number n of tips where Q($k_x$ in n) ≥ 99%. At this point the gambler might be faced with another challenge, when the mathematical software is not sufficient to produce reliable results for large faculties. A straightforward analytical solution for n does exist for the highest number of 13 strikes for any given safety level. With $k_{13}$ = 1, $M_{13}$ = 1 inserted into equation 3 one has

$$Q(k_{13} = 1 \text{ in n}) = \frac{\binom{1}{1}\binom{N-1}{n-1}}{\binom{N}{n}} = \frac{n}{N} \;, \tag{8}$$

which, solved for n, leads to

$$n = N\, Q(k_{13} = 1 \text{ in n}) \;. \tag{9}$$

Equation 9 contains what was shown in equation 6: absolute certainty, Q($k_{13}$ = 1 in n) = 100%, requires that the number n of tips equals the total amount N of combinations. If the gambler takes some risk and reduces the probability to Q($k_{13}$ = 1 in n) = 90%, the number n of tips required to achieve the highest number of 13 strikes reduces to n = 0,9 N, which provides a number still too large. A more moderate approach would be to determine the number of tips required to achieve at least one or more combinations with 10 strikes or more strikes for a given level of safety. An approximation is developed in the next chapter.

### 3.3. Probability to draw at least $k_x$ combinations with x strikes or more strikes in n tips

What does one or more combinations with 10 strikes or more strikes mean? In the jar containing in total N combinations, there are $M_{10}$ combinations with 10 strikes. Assuming a random guess with a hit rate of p = 1/3, N = $(1/p)^{13}$, equations 1 and 5 yield $M_{10}$ = 2.288 combinations. One or more combinations with 10 strikes are 1 or 2 or 3 or 4 …. or finally all 2.288 combinations with 10 strikes. More than 10 strikes are one or more combinations with 11 strikes or more strikes, one or more combinations with 12 strikes or more strikes, or the final combination with 13 strikes. A solution can be developed with the reverse probability. At first, we ask for the probability that we draw no combination with 10 strikes, no combination with 11 strikes, no combination with 12 strikes and no combination with 13 strikes in n tips. With equation 7 we get

$$Q(k_{10} = 0, k_{11} = 0, k_{12} = 0, k_{13} = 0 \text{ in n}) = \frac{\binom{M_{10}}{0}\binom{M_{11}}{0}\binom{M_{12}}{0}\binom{M_{13}}{0}\binom{N-M_{10}-M_{11}-M_{12}-M_{13}}{n}}{\binom{N}{n}} \,. \qquad (10)$$

The sum on the right hand side of equation 10 in abbreviated as $\sum_x M_x$ . $Q(k_{10} = 0, k_{11} = 0, k_{12} = 0, k_{13} = 0 \text{ in n})$ is abbreviated as $Q(k_x = 0 \text{ in n})$, x = 10, 11, 12, 13, which yields

$$Q(k_x = 0 \text{ in n}) = \frac{\binom{N-\sum_x M_x}{n}}{\binom{N}{n}} \,, x = 10, 11, 12, 13 \,. \qquad (11)$$

For further calculation, the fraction on the right hand side in equation 11 can be converted into a more useful expression (for details on the conversion, see appendix in chapter 6):

$$\frac{\binom{N-\sum_x M_x}{n}}{\binom{N}{n}} = \frac{\binom{N-n}{\sum_x M_x}}{\binom{N}{\sum_x M_x}} \,. \qquad (12)$$

With the constraint $\sum_x M_x \ll (N - n)$, x = 10, 11, 12, 13, the fraction on the right hand side in equation 12 can be approximated, so that equation 11 can be provided in the convenient form (for details on the approximation, see appendix in chapter 6)

$$Q(k_x = 0 \text{ in n}) = (1 - \frac{n}{N})^{\sum_x M_x} \,, x = 10, 11, 12, 13 \,. \qquad (13)$$

Equation 13 provides the probability to draw no combinations at all with 10, 11, 12, and 13 strikes in n tips. This is further elaborated into a measure for the level of safety to achieve at least a minimum number of strikes. To draw at least one or more combinations with 10 strikes or more strikes, the reverse probability gives

$$Q(k_x = 1 \text{ or more in n}) = 1 - (1 - \frac{n}{N})^{\sum_x M_x} \,, x = 10, 11, 12, 13 \,. \qquad (14)$$

Equation 14 can be solved for n. With $Q(k_x = 1 \text{ or more in n})$ abbreviated as Q we get

$$n = N\left[1 - (1 - Q)^{1/\sum_x M_x}\right] \,, x = 10, 11, 12, 13 \,. \qquad (15)$$

The probability Q is defined as a level of safety to achieve a minimum number of strikes. If we go further and ask for the number n of tips required to draw at least one or more combinations with 11 strikes or more strikes, equation 15 is applied with x = 11, 12, 13. For one or more combinations with 12 strikes or more strikes, x = 12, 13, and finally for 13 strikes, x = 13. For 13 strikes and $M_{13} = 1$, equation 15 reduces to the simple form provided in equation 9. Table 2 lists the number n of tips required to achieve one or more combinations with x strikes or more strikes for various safety levels Q, with the random guess hit rate p = 1/3.

| at least one or more combinations with x strikes or more strikes | Hit rate p = 33,33% Level of safety Q(kx = 1 or more in n) | | | |
|---|---|---|---|---|
| | 90% | 99% | 99,9% | 99,99% |
| 10 | 1.397 | 2.793 | 4.187 | 5.580 |
| 11 | 10.793 | 21.512 | 32.159 | 42.734 |
| 12 | 130.330 | 250.005 | 359.897 | 460.806 |
| 13 | 1.434.891 | 1.578.380 | 1.592.729 | 1.594.164 |

**Table 2**: Number n of tips required to achieve at least one or more combinations with x strikes or more strikes for various safety levels Q, hit rate p = 1/3.

The concept of the safety level introduced in equations 14 and 15 can be easily understood with table 2. In the first row, the low safety level of 90% requires to fill out just n = 1.397 tips to achieve at least one or more combinations with 10 strikes or more strikes. It follows that the opposite quantity, which could be appropriately named uncertainty U = (1 - Q), implies that there is still an uncertainty of 10% that the strategy fails, with all strikes among the remaining (N - n) = 1.592.926 combinations in the jar. In the last row, Q = 99,99% requires to fill out n = 1.594.164 tips to cover the combination with 13 strikes with a high safety. The uncertainty is quite low. With an uncertainty of U = 0,01% the combination with 13 strikes might still be among the remaining (N - n) = 159 combinations in the jar. The level of safety Q and the related uncertainty U should not be mixed up with the concept of risk applied in economic theory, which in the most common set-up, known as "Sharpe ratio", refers to the variance of profit. With a hit rate p = 1/3, gamblers on the average will remain on the losing side, except a few lucky ones. Even with a level of safety as low as Q = 90%, the required 1.397 tips to achieve at least one combination with 10 strikes amount to a total fee of 757€, with by far exceeds the expected return of 49€ (see table 1). For 11, 12, and 13 strikes, the negative business case for the gambler gets even worse.

### 3.4. Increasing the hit rate p, reducing the number of required tips

With a higher hit rate such as p = 1/2, the probability to achieve 13 strikes on a single tip increases to $P(13) = (1/2)^{13} = 1/8.192$, and the probability for 10 strikes increases to P(10) = 1/29 ($P(10) = M_{10}/N = 286/8.192$). The fraction for P(13) provided above reveals N = 8.192 combinations. The result might go against intuition, because 3 possibilities (win, draw, loss) produce $3^{13}$ = 1.594.323 combinations in all. The following remarks help to clarify the seeming discrepancy: First of all, when the hit rate p increases, the probability to achieve high numbers of strikes per tip should increase accordingly, as equation 1 proves, including P(13). 13 strikes appear in a single combination only, and P(13) = 1/N does only increase when N decreases. Secondly, assuming an upper (unrealistic) hit rate of p = 100%, which implies that a visionary gambler has the gift to correctly forecast any match, he would need to submit just a single tip to achieve all 13 strikes. From equation 1, it follows that P(13) = 1, and with $M_{13}$ = 1 equation 5 yields N = 1. For the visionary gambler, the total number of combinations N reduces to 1, the one and only combination. The context can be clarified with combinatorics, too. For a random guess, p = 1/3, the total number of combinations, N = $3^{13}$, does represent the maximum nescience, whereas for hit rates $p > 1/3$, there is prior knowledge which allows to reduce the number of combinations one has to pass through. This work further elaborates on p = 1/2, which is equivalent to a simple game with 2 mutually exclusive choices, each choice with equal chance. With 13 matches, the number of combinations one has to pass through reduces to N = $2^{13}$ = 8.192. Higher hit rates p are easy to obtain. Table 3 lists examples, including the random guess for reference.

| Method | Hit rate p | Reference, source |
|---|---|---|
| random guess | 33,33% | |
| tipster advice | ~ 43% | [12, 16] |
| better team wins* | 50,98% | Bundesliga season 2021/22 |
| home win | 50,42% | [12] |
| home win | 50,36% | Bundesliga season 1963/64 - 2021/22 |

**Table 3**: Hit rates for soccer provided by method, including the random guess (*better team according to rank in the final soccer table of the previous season).

More sophisticated methods do achieve even higher hit rates. Spann et al. [12] apply prediction market methods and betting odds, achieving hit rates p = 52,69% (prediction market) and p = 52,93% (betting odds). With rule-based forecasting and a combination of several methods applied to overlapping quantities of soccer games, they achieve p = 53,98% (prediction market and betting odds agree), p = 56,58% (prediction market and tipster agree), p = 56,52% (betting odds and tipster agree), and p = 57,11% (prediction markets, betting odds and tipster agree). Methods for obtaining hit rates beyond 51% are quite challenging to implement for the average gambler, they can be applied only for matches where reliable information is available at the same time for the same matches.

Compared to the aforementioned high-level techniques, the simple rules to be applied to achieve hit rates $p > 1/3$ listed in table 3 are more convenient to implement for the soccer betting TOTO 13er Wette. First of all, and next to discarding random guessing, the gambler should distrust any gut feelings biasing his judgements, especially when matches of his favorite soccer team are concerned. An inquiry in related sports indicates that gamblers tend to act more like fans than like investors [15]. Secondly, recommendations from soccer tipsters should be scrutinized. Soccer tipsters are independent experts publishing their forecasts online or in print journals. Tipsters don´t seem to apply formal statistical techniques to forecasting, relying instead on individual experience and intuition [16]. Even designated soccer experts - including sport journalists and soccer coaches - fail to predict matches more accurately than laymen [17]. Their low accuracy in forecasting, despite their high levels of expertise, indicates the inefficient use of information resulting from overconfidence, also found in the economic domain with the poor forecasting abilities of market pundits [4, 5]. The next method "better team wins" in table 3 does yield hit rates close to 51%, but this figure should be handled with care, because it was derived with a limited data base (a single season in the German premier league Bundesliga does account for 306 soccer matches only). The last two rows in table 3 refer to quite large and more reliable data sets from various soccer seasons. Obviously, home wins are the most frequent results of soccer matches. For further calculations, "home win" is applied with a hit rate $p = 1/2$, which is slightly lower than what can be achieved, but which already serves the purpose of this work. If we take $p = 1/2$, calculate the probabilities for x strikes $P(x)$, $x = 10, 11, 12, 13$ with equation 1, the number of combinations $M_x$ for x strikes with equation 5, and apply equation 15 to derive the number n of tips required for various levels of safety, we obtain table 4:

| at least one or more combinations with x strikes or more strikes | Hit rate p = 50%<br>Level of safety Q(kx = 1 or more in n) | | | |
|---|---|---|---|---|
| | 90% | 99% | 99,9% | 99,99% |
| 10 | 50 | 100 | 149 | 198 |
| 11 | 203 | 400 | 593 | 781 |
| 12 | 1.243 | 2.297 | 3.191 | 3.949 |
| 13 | 7.373 | 8.110 | 8.184 | 8.191 |

**Table 4**: Number n of tips required to achieve at least one or more combinations with x strikes or more strikes for various safety levels Q, hit rate $p = 1/2$.

# 4. Results

## 4.1. Traits of inefficiency in the German soccer betting TOTO 13er Wette

Tables 5a, 5b and 5c and show the expected profit in € for the hit rates 43%, 47% and 50%. Each row refers to the number of strikes x = 10, 11, 12, 13, for which the lottery agency distributes returns. The total fees are calculated based on the number n of tips required to achieve at least x strikes (0,50€ per tip plus 0,50€ per voucher, each voucher containing 12 tips). The expected profit calculates to

$$\text{expected profit} = \text{expected return} - \text{total fees} \ , \qquad (16)$$

it is listed in the columns on the right hand side, with positive profit highlighted with green cells, and negative profit (loss) highlighted with red cells. Total fee, number n of required tips, and expected profit are split into 4 levels of safety Q. We need to highlight two important things: first of all, the expected return, and in consequence the expected profit, are average quantities. The actual weekly returns distributed by the lottery agency may differ from the expected returns listed in the tables 5a to 5c, resulting in variances around average values. As briefly noted at the end of chapter 3.3, variances around expected profits could be treated with additional concepts of risk from economic theory ("Sharpe ratio"). For the purpose of this work, which is about demonstrating inefficiencies of soccer betting, averages are sufficient. On the whole, the figures provide adequate guidance for the returns and resulting profits to be expected. Uncertainties regarding actual returns are compensated by a beneficial factor: for each row in tables 5a to 5c, the quantity n and the resulting fees refer to the condition to achieve at least one or more combinations with x strikes, or more strikes. This implies that the gambler might end up with more than just a single combination with x strikes, or with even more strikes, for which the distributed returns are higher.

Table 5a again illustrates why tipster recommendations for match results should be scrutinized. For any acceptable level of safety Q ≥ 99%, the business case for the gambler is negative. Despite tremendous costs and efforts, he will lose. The business case appears to be positive for the highest level of 13 strikes only, for which the gambler would need to fill out numerous tips with total fees worth half his annual salary. Since the actual return fluctuates around the average, as noted in the previous passage, the strategy to follow tipster advice should be discarded as too risky.

| a) Hit rate p = 43% | | Total fee (number n of required tips) | | | | Expected profit (number n of required tips) | | | |
|---|---|---|---|---|---|---|---|---|---|
| at least one or more combinations with x strikes or more strikes | Expected return | Level of safety Q(kx = 1 or more in n) | | | | | | | |
| | | 90% | 99% | 99,9% | 99,99% | 90% | 99% | 99,9% | 99,99% |
| 10 | 49 € | 89 € | 177 € | 265 € | 352 € | -40 € | -128 € | -216 € | -303 € |
| | | (n=163) | (n=326) | (n=488) | (n=649) | (n=163) | (n=326) | (n=488) | (n=649) |
| 11 | 434 € | 465 € | 922 € | 1.373 € | 1.816 € | -31 € | -488 € | -939 € | -1.382 € |
| | | (n=857) | (n=1.701) | (n=2.533) | (n=3.352) | (n=857) | (n=1.701) | (n=2.533) | (n=3.352) |
| 12 | 6.449 € | 3.741 € | 7.037 € | 9.942 € | 12.503 € | 2.708 € | -588 € | -3.493 € | -6.054 € |
| | | (n=6.905) | (n=12.991) | (n=18.354) | (n=23.081) | (n=6.905) | (n=12.991) | (n=18.354) | (n=23.081) |
| 13 | 107.813 € | 28.372 € | 31.210 € | 31.494 € | 31.522 € | 79.441 € | 76.604 € | 76.320 € | 76.291 € |
| | | (n=52.379) | (n=57.617) | (n=58.141) | (n=58.193) | (n=52.379) | (n=57.617) | (n=58.141) | (n=58.193) |

| b) Hit rate p = 47% | | Total fee (number n of required tips) | | | | Expected profit (number n of required tips) | | | |
|---|---|---|---|---|---|---|---|---|---|
| at least one or more combinations with x strikes or more strikes | Expected return | Level of safety Q(kx = 1 or more in n) | | | | | | | |
| | | 90% | 99% | 99,9% | 99,99% | 90% | 99% | 99,9% | 99,99% |
| 10 | 49 € | 44 € | 87 € | 130 € | 173 € | 5 € | -38 € | -81 € | -124 € |
| | | (n=81) | (n=160) | (n=240) | (n=319) | (n=81) | (n=160) | (n=240) | (n=319) |
| 11 | 434 € | 198 € | 390 € | 580 € | 765 € | 236 € | 44 € | -146 € | -331 € |
| | | (n=364) | (n=720) | (n=1.069) | (n=1.412) | (n=364) | (n=720) | (n=1.069) | (n=1.412) |
| 12 | 6.449 € | 1.357 € | 2.528 € | 3.539 € | 4.411 € | 5.092 € | 3.921 € | 2.910 € | 2.038 € |
| | | (n=2.504) | (n=4.666) | (n=6.532) | (n=8.143) | (n=2.504) | (n=4.666) | (n=6.532) | (n=8.143) |
| 13 | 107.813 € | 8.928 € | 9.820 € | 9.910 € | 9.918 € | 98.886 € | 97.993 € | 97.904 € | 97.895 € |
| | | (n=16.481) | (n=18.129) | (n=18.294) | (n=18.310) | (n=16.481) | (n=18.129) | (n=18.294) | (n=18.310) |

| c) Hit rate p = 50% | | Total fee (number n of required tips) | | | | Expected profit (number n of required tips) | | | |
|---|---|---|---|---|---|---|---|---|---|
| at least one or more combinations with x strikes or more strikes | Expected return | Level of safety Q(kx = 1 or more in n) | | | | | | | |
| | | 90% | 99% | 99,9% | 99,99% | 90% | 99% | 99,9% | 99,99% |
| 10 | 49 € | 28 € | 55 € | 81 € | 108 € | 21 € | -6 € | -32 € | -59 € |
| | | (n=50) | (n=100) | (n=149) | (n=198) | (n=50) | (n=100) | (n=149) | (n=198) |
| 11 | 434 € | 110 € | 217 € | 322 € | 424 € | 324 € | 217 € | 112 € | 10 € |
| | | (n=203) | (n=400) | (n=593) | (n=781) | (n=203) | (n=400) | (n=593) | (n=781) |
| 12 | 6.449 € | 674 € | 1.245 € | 1.729 € | 2.140 € | 5.775 € | 5.204 € | 4.720 € | 4.309 € |
| | | (n=1.243) | (n=2.297) | (n=3.191) | (n=3.949) | (n=1.243) | (n=2.297) | (n=3.191) | (n=3.949) |
| 13 | 107.813 € | 3.994 € | 4.393 € | 4.433 € | 4.437 € | 103.819 € | 103.420 € | 103.380 € | 103.376 € |
| | | (n=7.373) | (n=8.110) | (n=8.184) | (n=8.191) | (n=7.373) | (n=8.110) | (n=8.184) | (n=8.191) |

**Tables 5a, 5b, 5c**: Expected profit in € for the hit rates a) 43%, b) 47% and c) 50%. The rows refer to the number of strikes x = 10, 11, 12, 13, for which the lottery agency distributes returns. The total fee is calculated based on the number n of required tips. Positive profit is highlighted with green cells, and negative profit (loss) is highlighted with red cells. Total fee, number of tips, and expected profit are split into 4 levels of safety Q. All quantities in € are rounded up to integer numbers.

Table 5b is introduced to show the sensitivity of the expected profits with changes of the hit rate. From p = 43% in table 5a to p = 47% in table 5b, the relative change of the hit rate is $\Delta p/p \sim 9\%$ only, but the business case becomes positive for acceptable levels of safety $Q \geq 99\%$ and related manageable fees.

Table 5c refers to the method "home win" with p = 50%. It shows positive profit for x = 11, 12, and 13 strikes and for all levels of safety. Assuming $Q \geq 99\%$ as acceptable level of safety to spend a few hundred € total fees on a weekly basis for achieving at least 11 strikes, the resulting profit is positive, but with values ranging from 324€ down to 10€ a quite meager extra income. A few thousand € would need to be spent on fees for achieving at least 12 strikes or at least 13 strikes, where the expected profit would equal monthly incomes (x = 12 strikes) or even an annual incomes (x = 13 strikes). With a few thousand € total fees, one should minimize risk and choose higher safety levels. The ultimate measure would be to choose Q = 100%. For 13 strikes one has to fill out 8.192 tips (see equation 9), which is simply 1 tip more than required for the safety level Q = 99,99% (see table 5c). The combinations would have to be generated on 8.192 tips in such a way that home team wins are checked equally often as draws and away team wins combined. Over all tips, this method results in 50% home wins, the residual 50% distributed between draws and away team wins (home team losses). The challenges are to (i) stay with the 50% home team wins and (ii) avoid any doubles when filling out thousands of tips, a task better consigned to a software program. Basically, one could make a living with filing tips containing thousands of combinations to the lottery agency – if there weren´t any further drawbacks.

### 4.2. Barriers hampering high profits: transaction efforts and betting regulations

There are two basic barriers for filing numerous tips to the lottery agencies. The first one refers to transaction efforts. As noted at the end of chapter 2.1, 8 out of 16 state-run lottery agencies do accept the filled out vouchers for the TOTO 13er Wette offline on paper only, to be filed to local lottery shops. Assuming a gambler intends to hand in 4.000 tips, he needs to secure 334 paper vouchers first (see voucher in figure 1a for reference). This might exceed the total volume of vouchers on stock in the lottery shop. Even if the gambler does virtually loot several lottery shops to accumulate a sufficient amount of paper vouchers, the next challenge is how to fill out the tips with various combinations. Machines aren´t available to accomplish this task, so the gambler needs to sit down with a ball pen and fill out the vouchers himself. Doubtless, the procedure is beyond human capacity and patience. On lower scales, the expected profits put associated efforts into question. If the gambler somehow manages to fill out a few hundred tips, say 600 tips on 50 vouchers, it might take a few days of focused work. With 600 tips, one could achieve at least 11 strikes with a safety level of 99,9% (see table 5c). In case the gambler is in need, the expected profit of 112€ might justify aforementioned efforts. It is self-explanatory that the transaction costs of filing tips on paper vouchers are extremely high.

Efforts drop marginally with online submissions of vouchers to the lottery agencies, which are possible in 8 federal states (see online voucher in figure 1b for reference). After having registered with an account at the lottery agency, it turns out that the online procedure is designed in such a way that it resembles the offline procedure: a pop-up menu representing a voucher opens, where the gambler needs to check his guesses for 13 matches in each tip with his cursor. Filling out hundreds or even thousands of tips equals the offline paper process, the task is just transferred from the ball pen to the cursor. Even for software adept gamblers, there is no obvious starting point on how to automatize the process on user side. There is one noteworthy exception for online submissions: in the federal state of Nordrhein-Westfalen, the lottery agency allows to upload data files containing tips with combinations generated beforehand by the user. The required data format csv (comma separated values) is widespread and can be generated and processed with simple text editors. The maximum number of tips per file for upload is 1.250. When the gambler wants to submit more tips, he can do that with additional files. The lottery agency supports some specific TOTO-programs prevalent among soccer enthusiasts, which convert the tips into the data structure and the file format required for upload (for details, see sub-menu csv-upload on https://www.westlotto.de/service/hilfe-faqs/hilfe_faq.html), but there seems no basic hindrance to accomplish this task oneself. For the latter case, the efforts reduce to writing a software program which generates the tip combinations beforehand. In addition to that, gamblers would be required to relocate to the federal state of Nordrhein-Westfalen and open their online account from there, in case they have their permanent residence elsewhere.

The second basic barrier is imposed by betting regulations. In all federal states, there are financial limits to spend on fees for online gambling. The majority of states puts the upper limit to 1.000€ per month, which can be spent by the gambler. At the time of writing of this work, two states deviate with maximum limits of 1.100€ and one state with 1.500€ per month. A third state limits the account to 300€ per week (which adds up 1.200€ per month, but limits each weekly bet to just 552 tips). If all fees exceeding 1.000€ are crossed out in table 5c, we end up with table 6. The number of tips which can be submitted online is diminished considerably. In the row showing 12 strikes or more, there is an option left with 1.243 tips, 674€ total fee, and a related expected profit with 5.775€. However, this should be handled with care, because it refers to a safety level of 90% only. The safety level implies an uncertainty of 10% that 12 strikes or more are still among the remaining (N – n) = 6.949 combinations. In the latter case, an unlucky fellow will end up in the row with 11 strikes or more, for which he receives a return of 434€, ending up with a loss of 434€ - 674€ = - 240€. For more courageous gamblers, who want to take up with a safety level of 90%, the following recipe to deal with the uncertainty U = 10% might work. The probability that a courageous gambler will fail 4 times in a row calculates to

$$P(4 \text{ x fail}) = U^4 = 0{,}01\,\% \,. \tag{17}$$

The reverse probability that a courageous gambler will succeed at least once is

$$P(\text{at least 1 x success}) = 1 - U^4 = 99{,}99\,\% \,. \tag{18}$$

Worst case, the gambler has to spend 4 x 674€ = 2.696€ to achieve the expected return of 6.449€. The expected profit of this strategy would be 6.449€ - 2.696€ = 3.753€. It depends on the gambler´s strength of nerve to pull through this venture, especially when he already failed 3 times in a row. Starting with a safety level of 90% promises an enthralling experience. For acceptable safety levels Q ≥ 99%, gamblers are confined to achieve just 11 strikes or more. In consequence, the expected profit is reduced to a few hundred € at best.

| Hit rate p = 50% | | Total fee (number n of required tips) | | | | Expected profit (number n of required tips) | | | |
|---|---|---|---|---|---|---|---|---|---|
| at least one or more combinations with x strikes or more strikes | Expected return | Level of safety Q(kx = 1 or more in n) | | | | | | | |
| | | 90% | 99% | 99,9% | 99,99% | 90% | 99% | 99,9% | 99,99% |
| 10 | 49 € | 28 € (n=50) | 55 € (n=100) | 81 € (n=149) | 108 € (n=198) | 21 € (n=50) | -6 € (n=100) | -32 € (n=149) | -59 € (n=198) |
| 11 | 434 € | 110 € (n=203) | 217 € (n=400) | 322 € (n=593) | 424 € (n=781) | 324 € (n=203) | 217 € (n=400) | 112 € (n=593) | 10 € (n=781) |
| 12 | 6.449 € | 674 € (n=1.243) | ~~1.245 €~~ (n=2.297) | ~~1.729 €~~ (n=3.191) | ~~2.140 €~~ (n=3.949) | 5.775 € (n=1.243) | ~~5.204 €~~ (n=2.297) | ~~4.720 €~~ (n=3.191) | ~~4.309 €~~ (n=3.949) |
| 13 | 107.813 € | ~~3.994 €~~ (n=7.373) | ~~4.393 €~~ (n=8.110) | ~~4.433 €~~ (n=8.184) | ~~4.437 €~~ (n=8.191) | ~~103.819 €~~ (n=7.373) | ~~103.420 €~~ (n=8.110) | ~~103.380 €~~ (n=8.184) | ~~103.376 €~~ (n=8.191) |

**Table 6:** Same as table **5c**, with all fees exceeding 1.000€ crossed out due to legal betting regulations. The related expected profits are crossed out, too.

Lottery agencies justify the upper limit of around 1.000€ per month with betting regulations. A major update of the underlying legal framework, named Glückspielstaatsvertrag, came into effect on 1st of July 2021 [18]. One of the main purposes is to prevent individuals from compulsive gambling and financial ruin. As a secondary effect, the regulations seem to impose upper limits for systematically generating profits large enough to make a living. However, there is some positive expected profit left in table 6, which demonstrates that the soccer betting TOTO 13er Wette shows traits of inefficiency. It depends on the gambler if the profits are a sufficient incentive to submit hundreds of tips, getting along with the financial limitations. The courageous gambler, who bets several times in a row with a safety level of Q = 90%, faces a similar challenge. Due to the upper fee of 1.000€ per month, betting 4 times in a row might cover 3 – 4 months. This long period is clouded by permanent uncertainty if the strategy finally works out with the next soccer match results, if the previous ones failed. Waiting up to 4 months for an expected profit of 3.753€ poses a tough ordeal.

## 5. Summary, discussion and outlook

This work provides a strategy to systematically generate profits with the soccer betting TOTO 13er Wette. High hit rates for the match outcomes are based on the observation that home wins are the most frequent results of soccer matches. A recipe is developed to determine the number of lottery tips required to achieve a specific number of strikes for any given level of safety. On the way, elaborated in the appendix, a useful approximation is derived to cope with large numbers in hypergeometric distributions, valid under certain constraints. The whole approach shifts soccer betting from a game of chance towards a game of skill. From a market theory point of view, the expected profits do reveal an anomaly, indicating traits of inefficiency for the soccer betting TOTO 13er Wette.

Due to seemingly out of time transaction procedures and restrictive betting regulations, the profits are melted down to moderate levels where associated efforts question the scheme. A rather uncommon strategy is briefly outlined for courageous gamblers, which might extend over several months to obtain expected profits of appreciable quantity. The procedure requires endurance and strength of nerve, because it is based upon the recipe that success is getting closer with the number of fails accomplished up to that point. Likewise observations from the soccer betting market are found elsewhere: Spann et al. [12] demonstrated that systematic profits are foiled by high fees charged by bookmakers. Kaunitz et al. [11] revealed inefficiencies in soccer betting by repeatedly beating commercial, privately owned bookmakers with their own numbers. Although the researchers played according to the rules, they experienced discriminatory practices from some bookmakers imposed upon their accounts, limitations which presumably serve to compensate for the market inefficiencies. Apparently, smart individuals and professional bettors were hindered to exploit any loopholes, which were retroactively closed by altering the gambling rules.

The financial market served to introduce the concept of the efficient market hypothesis. The question is, if lessons learned from soccer betting can be transferred back to financial markets. Mandelbrot et al. [19] state that financial markets are not as efficient and balanced towards equilibrium as standard financial theory assumes. More often these days, the market is spiking up or dropping down immediately with trade opening, when the unconcerned shareholder still enjoys morning coffee. The dynamics of such bursts defies any concept of equilibrium. Contrasting common belief, inquiries demonstrate that only a few big players holding large stock volumes are responsible for the major – and therefore crucial - fluctuations of share prices [20, 21, 22]. Data analyzed from the New York Stock Exchange show that trades by retail investors represent less than 2% of the trading volume [23]. If prices are determined mainly by large institutional investors, it probably shifts market efficiency out of balance at a certain level. Research indicates that institutional investors utilize asymmetries such as information advantages, exploiting the price momentum at the expense of individual shareholders [24]. It is no coincidence that recent work took up with unexpected discontinuous structural market transitions and diversity breakdowns [25]. Are prices still formed by an equilibrium of supply and demand? Are players such as hedge funds, by virtue of large transactions and market power, exerting normative forces, which misalign the business to their favor [26]? The latter would effectively constitute an asymmetry similar to that one between bookmakers and bettors. Financial markets don´t seem to be fully efficient. Similar to sports betting, there should be levers for outperformance.

The uncertainties related to soccer match results hold some analogies to the randomness of stock prices, which is subject of further research. If nothing is known about a single stock, and the shareholder doesn´t care, picking stocks resembles a random guess between 3 future alternatives: the stock price can fall, it can stay at the current level, or it can rise. The hit rate would be $p = 1/3$, the selection method corresponds to a strategy of economist Burton Malkiel´s distinguished investor team, humorously called "the dart throwing monkeys" [27]. If the shareholder is indifferent to the portfolio containing a large mix of various stocks, selected without bias, its value should just mirror the market average, as predicted by market efficiency [28, 29]. Chapter 3 shows that the only chance to generate consistent positive profit above (negative) average in soccer betting is to improve the hit rate for single match outcomes. All other parameters and relations, which make up the major part of the underlying mathematical theory and constitute the system and the rules of soccer betting, can´t be altered to the gambler´s advantage. Transferred to the stock market, it means that retail investors would be left with improving their stock picking skills and arbitrage to exploit any market inefficiencies, as long as they aren´t in a position to change the rules or even to turn the tables [30]. More general questions refer to the distribution of profit, with the average on the negative side (loss) for gamblers in sports betting, and on the positive side for retail investors in stock markets, as long as there is sustainable economic growth witnessed for the past decades. If we subtract economic growth and inflation, would the average of the distribution of profit for retail investors center at zero? Or would it be negative similar to sports betting? The latter would indicate a drain to somewhere else, which leads to the question: who in finance assumes a role analog to bookmakers?

Chapter 3.4 illustrates that one shouldn´t follow pundit recommendations with blind trust, which equally applies to the sports and the financial domain [4, 5, 17]. A deep, but narrow focus can miss the bigger picture. Aggravated with expert fallacies and overconfidence biases, one might end up with misleading forecasts. As a rule of thumb, pundits spreading frequent advice in the news presumably don´t have a Midas touch, otherwise they wouldn´t have to work. Most likely, their hidden agenda is to establish themselves as gurus through presence in the media, and to trigger herding among unsophisticated followers [31]. More reputable professionals are well aware of the market caprice and keep a low profile, rather performing the "master of the obvious" in interviews.

This article and references herein highlight the importance of prior knowledge and skills. For most matters in life, this isn´t groundbreaking news – but within the framework of efficient market theory, it somewhat helps to re-establish the value of know-how [32, 33]. The author Ed Smith, who compares insights gained from sports with those from other spheres of life, writes: "Sports, like trading markets, occasionally throw up inefficient blind spots – areas where conventional wisdom underestimates the true value of one piece of the strategic jigsaw. These neglected corners of the game, like undervalued stock, provide a systematic opportunity" [34]. Although profits are moderate, this work shows an example that indeed, "one can beat the average".

## 6. Appendix

**(i)** The following identity has to be shown for equation 12:

$$\frac{\binom{N-\sum_x M_x}{n}}{\binom{N}{n}} = \frac{\binom{N-n}{\sum_x M_x}}{\binom{N}{\sum_x M_x}} \ .$$

The step by step conversion yields

$$\frac{\binom{N-\sum_x M_x}{n}}{\binom{N}{n}} = \frac{\frac{(N-\sum_x M_x)!}{n!\,(N-n-\sum_x M_x)!}}{\frac{N!}{n!\,(N-n)!}} = \frac{\frac{(N-\sum_x M_x)!}{(N-n-\sum_x M_x)!}}{\frac{N!}{(N-n)!}} = \frac{\frac{(N-n)!}{(N-n-\sum_x M_x)!}}{\frac{N!}{(N-\sum_x M_x)!}} = \frac{\frac{(N-n)!}{(\sum_x M_x)!\,(N-n-\sum_x M_x)!}}{\frac{N!}{(\sum_x M_x)!\,(N-\sum_x M_x)!}} = \frac{\binom{N-n}{\sum_x M_x}}{\binom{N}{\sum_x M_x}} \ . \qquad \text{q.e.d.}$$

**(ii)** It has to be shown that for $\sum_x M_x \ll (N-n)$, $x = 10, 11, 12, 13$ the following approximation can be applied in equation 13:

$$\frac{\binom{N-n}{\sum_x M_x}}{\binom{N}{\sum_x M_x}} \approx (1-\frac{n}{N})^{\sum_x M_x} \ .$$

At first it is shown that for A, b, n ∈ N and b << A, which can be expressed as A ⟶ ∞, the following approximation holds:

$$\frac{\frac{A!}{(A-b)!}}{\frac{(A+n)!}{(A+n-b)!}} \approx \left(\frac{A}{A+n}\right)^b \ .$$

Numerator and denominator of the fraction have the same number of factors, for A ⟶ ∞ we get the threshold value

$$\lim_{A\to\infty} \frac{\frac{A!}{(A-b)!A^b}}{\frac{(A+n)!}{(A+n-b)!(A+n)^b}} = \lim_{A\to\infty} \frac{\left(1-\frac{1}{A}\right)\left(1-\frac{2}{A}\right)\ldots\left(1+\frac{1}{A}-\frac{b}{A}\right)}{\left(1-\frac{1}{A+n}\right)\left(1-\frac{2}{A+n}\right)\ldots\left(1+\frac{1}{A+n}-\frac{b}{A+n}\right)} = 1 \ .$$

It follows that

$$\frac{\frac{A!}{(A-b)!A^b}}{\frac{(A+n)!}{(A+n-b)!(A+n)^b}} \approx 1, \text{ which after conversion gives } \frac{\frac{A!}{(A-b)!}}{\frac{(A+n)!}{(A+n-b)!}} \approx \left(\frac{A}{A+n}\right)^b \ .$$

With the substitutions $A = (N - n)$ and $b = \sum_x M_x$ we get

$$\frac{\frac{(N-n)!}{(N-n-\sum_x M_x)!}}{\frac{N!}{(N-\sum_x M_x)!}} \approx \left(1-\frac{n}{N}\right)^{\sum_x M_x}, \text{ which gives } \frac{\binom{N-n}{\sum_x M_x}}{\binom{N}{\sum_x M_x}} \approx \left(1-\frac{n}{N}\right)^{\sum_x M_x} \ . \qquad \text{q.e.d.}$$